%% file: ijcai21.tex
\documentclass{article}
\usepackage{ijcai21}
\usepackage{lmodern,babel,adjustbox,booktabs,multirow}
\usepackage{times}
\usepackage{soul}
\usepackage{url}
\usepackage[hidelinks]{hyperref}
\usepackage[utf8]{inputenc}
\usepackage{graphicx}
\usepackage{amsmath}
\usepackage{amsthm}
\usepackage{booktabs}
\usepackage{algorithm}
\input{macro}

\usepackage{appendix}
\usepackage{stfloats}

\title{Point-based Acoustic Scattering for Interactive Sound Propagation via Surface Encoding
}
\author{
Hsien-Yu Meng\and
Zhenyu Tang \and
Dinesh Manocha\\
\affiliations
University of Maryland, College Park
\emails
\{mengxy19,zhy,dmanocha\}@umd.edu
}

\begin{document}

\maketitle
\begin{abstract}
    
We present a novel geometric deep learning method to compute the acoustic scattering properties of geometric objects. Our learning algorithm uses a point cloud representation of objects to compute the scattering properties and integrates them with ray tracing for interactive sound propagation in dynamic scenes. We use discrete Laplacian-based surface encoders and approximate the neighborhood of each point using a shared multi-layer perceptron. We show that our formulation is permutation invariant and present a neural network that computes the scattering function using spherical harmonics.
Our approach can handle objects with arbitrary topologies and deforming models, and takes
less than 1ms per object on a commodity GPU. We have analyzed the accuracy and perform validation on thousands of unseen 3D objects and highlight the benefits over other point-based geometric deep learning methods. To the best of our knowledge, this is the first real-time learning algorithm that can approximate the acoustic scattering properties of arbitrary objects with high accuracy.
\end{abstract}

\input{1}

\input{2}
\input{3}
\input{4}

\input{5}
\input{6}

\section*{Acknowledgments}
This work was supported in part by ARO grants W911NF-18-1-0313 and W911NF-19-1-0315, NSF grants \#1910940, Adobe, Facebook and Intel.
\bibliographystyle{named}
\bibliography{ijcai21}
\pagebreak
\clearpage

\appendix
\appendixpage
\input{appendix}

\end{document}

%% file: macro.tex
\usepackage{color}
\makeatletter
\def\zz{\edef\zzz{\pdfliteral{\current@color}}}
\makeatother

\makeatletter
\newcommand\footnoteref[1]{\protected@xdef\@thefnmark{\ref{#1}}\@footnotemark}
\makeatother

\newcommand{\shorteq}{%
  \settowidth{\@tempdima}{-}
  \resizebox{\@tempdima}{\height}{=}%
}

\usepackage{amsmath}
\usepackage{subcaption}
\usepackage[utf8]{inputenc}
\usepackage[linewidth=1pt]{mdframed}
\usepackage[T1,OT1]{fontenc}
\usepackage{babel}
\usepackage[font=small,belowskip=-1.5pt]{caption} 
\usepackage[noend]{algpseudocode}
\usepackage{array}
\usepackage{graphicx}
\usepackage{amsfonts}
\usepackage{enumitem}
\usepackage{soul}
\usepackage{hhline}
\usepackage{multirow, makecell}
\usepackage{float}
\usepackage{booktabs}

\usepackage{amsthm}
\usepackage{color}
\usepackage{transparent}
\usepackage{url}
\usepackage{setspace}
\usepackage{mathtools}
\theoremstyle{plain}

\newtheorem{theorem}{Theorem}[section]

\newtheorem{lemma}[theorem]{Lemma}

\usepackage{amssymb}
\usepackage{pifont}
\newcommand{\xmark}{\ding{55}}%

%% file: 1.tex

\section{Introduction}


Many interactive applications like games and virtual environments need to generate realistic or plausible sound effects. The resulting audio-visual renderings can enhance the realism and improve a user's level of immersion~\cite{liu2020sound}. Current techniques used in game engines or VR systems for interactive sound rendering are based on reverberation filters or ray tracing. However, these methods are unable to generate realistic low-frequency sound effects corresponding to diffraction or interference.

The most accurate methods for sound propagation and rendering directly solve the acoustic wave-equation using numeric solvers. These techniques either compute the global acoustic pressure field for the entire scene~\cite{raghuvanshi2018parametric} or compute the {\em acoustic scattering effects} of each separated object~\cite{rungta2018diffraction,kuttruff2016room}. The global methods are very expensive, have a high memory overhead, and are limited to static scenes. 

Acoustic scattering corresponds to the disturbance of a given incident sound field due to an object's shape and surface properties. It can be regarded as one of the fundamental characteristics of an object. The effect of scattering can be expressed in terms of a scattered sound field. 
The computation of acoustic scattering fields is also useful for other applications, including sound synthesis, sound source localization, object recognition, etc. The most accurate methods for acoustic scattering field computation use solvers like  boundary-element methods. In practice, accurate computation of acoustic scattering field of each object can take a few minutes and cannot be directly used within an interactive application. Instead, some hybrid sound rendering systems precompute the scattering fields of rigid objects and combine them with interactive ray tracing at runtime.
Recently, some machine learning based methods have been proposed to compute the scattering fields~\cite{pulkki2019machine,fan2019fast,tang2020learning}. However, they are limited to simple shapes (e.g., convex shapes) or only work well for low frequencies.

\noindent {\bf Main Results:} We present novel techniques based on geometric deep learning on differential coordinates to approximate the acoustic scattering properties of arbitrary objects. Our approach is general and makes no assumption about object's shape, genus, or rigidity. We approximate the objects using point clouds, and each point in the point cloud representation is encoded in a high-dimensional latent space. Moreover, the local surface shapes in the latent space are encoded using surface encoders. This enables us to handle arbitrary topology. Our network takes the point cloud as an input and outputs the spherical harmonic coefficients that represent the acoustic scattering field. 

We present techniques to generate the ground truth data using an accurate wave-solver on a large geometric dataset. We have evaluated the performance on thousands of objects that are very different from the training database (with varying convexity, genus, shape, size, orientation) and observe high accuracy. We also perform an ablation study to highlight the benefits of our approach. 
The additional runtime overhead of estimating the scattering field from neural networks is less than $1$ms per object on a NVIDIA GeForce RTX 2080 Ti GPU. We also prove that our learning method is permutation and rotation invariant, which is an important characteristic for accurate computation of acoustic scattering fields. We have also integrated our geometric learning algorithm with ray tracing for hybrid sound propagation in interactive applications.


%% file: 2.tex


\section{Related Works}

\subsection{Wave-Acoustic Simulation}
Wave-acoustic simulation  methods aim to solve the wave equation that governs the propagation and scattering of sound waves. 
Some conventionally used numeric methods include the finite-element method, the boundary-element method and the finite-difference time domain, as surveyed in ~\cite{liu2020sound} 
The common requirement of these methods is the proper discretization of the problem domain (e.g., spatial resolution, time resolution), which means the time complexity of them will scale drastically with the simulation frequency. 
An alternative way to use wave simulation results in real-time applications is to pre-compute a large amount of sound fields in a scene and compressing the results using perceptual metrics~\cite{raghuvanshi2018parametric}. However, they are limited to static scenes.

\subsection{Learning-Based Acoustics}
There is considerable work on developing machine learning methods for applications corresponding to audio-visual analysis and acoustic scene classification~\cite{bianco2019machine}. In comparison, there are much fewer works studying the generation of acoustic data from a physical perspective. 
\cite{pulkki2019machine} propose to use a neural network to model the acoustic scattering effect from rectangular plates.
Recently, \cite{fan2019fast} train convolutional neural networks (CNNs) to learn to map planar sound fields induced by convex scatterers in 2D. \cite{tang2020learning} uses the well-known PointNet method to compute the scattering properties for point clouds, but it only works well for low frequencies.
Motivated by the last method, we aim to develop networks that can deal with object geometry in 3D and extend the generality of this learning-based approach. 


\subsection{Geometric Deep Learning and Shape Representation}

There is considerable recent work on generating plausible shape representations for 3D data, including voxel-based~\cite{voxelnet}, point-based~\cite{pointnet,qi2017pointnet++,dgcnn} and mesh-based~\cite{hanocka2019meshcnn} shape representations. 
This includes work on shape representation by learning implicit surfaces on point clouds~\cite{smirnov2019deep}, 
hierarchical graph convolution on meshes~\cite{structureNet}, 
encoding signed distance functions for surface reconstruction~\cite{deepsdf}, etc. 
However, previous methods on point cloud shape representations learn by designing loss functions to constrain surface smoothness on global Cartesian coordinates. Such functions only provide spatial information of each point and lack information about local shape of the surface compared to the explicit 
discretization of the continuous Laplace-Beltrami operator and curvilinear integral.
Instead, we use point cloud based learning algorithms, which do not require mesh Laplacians for graph neural networks. This makes our approach applicable to all kind of dynamic objects, including changing topologies. 



%% file: 3.tex
\section{Our Approach}

\begin{figure}[htbp]
  \centering
  \includegraphics[width=\linewidth]{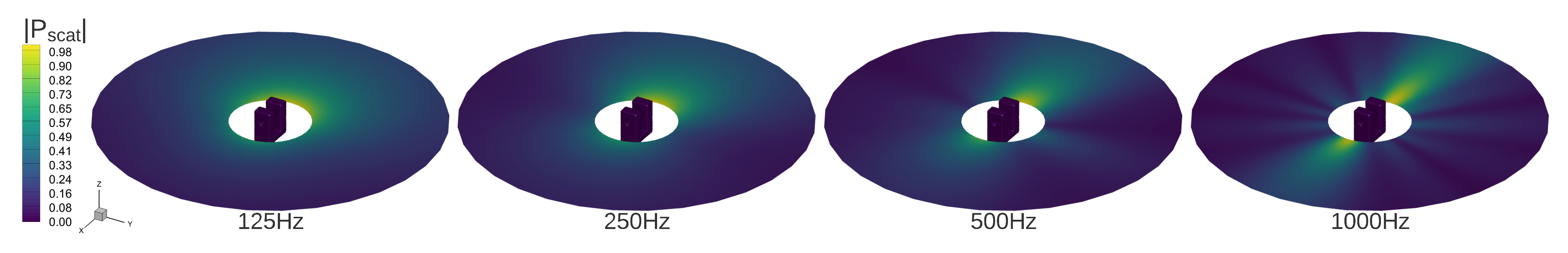}
  \caption{Acoustic scattering of the same object at four different frequencies, assuming a plane wave travelling to the $-x$ direction and a sound-hard boundary condition for the object/scatterer in the center. The scattering patterns are significantly different among frequency bands. Our learning method can compute the scattering fields with high accuracy for arbitrary objects at interactive rates, including at higher frequencies.}
  \label{fig:scattering}
\end{figure}

Our goal is to directly predict the acoustic scattering field (ASF) resulting from an incident wave and a sound scatterer of known shape (described by a triangle mesh). 
A set of frequency dependent ASFs are visualized in Figure~\ref{fig:scattering} on a disk plane. 
The two key steps are: (i) defining a compact and efficient way to encode the ASF; (ii) designing a neural network architecture that can effectively exploit the 3D geometric information so as to establish the mapping between the object shape and the ASF. 
In this section, we provide necessary background on wave acoustics and present an overview of our approach.





\subsection{Acoustic Scattering Fields}
\label{sec:asf}

We are interested in knowing how the incident sound field will be scattered when it acts on an object surface. Let the sound pressure at location $\bf{x}$ and time $t$ be $P(\bf{x},t)$, and the wave equation describes the propagation of sound:
\begin{equation}
\label{eq:waveequation}
(\nabla ^2 - \frac{1}{c^2} \frac{ \partial^2 }{\partial t ^2 })P(\bf{x},t) = 0,
\end{equation}
where $c$ is the speed of sound, which is commonly assumed to be constant in non-large scale indoor environments. 
Transforming into the frequency domain, we obtain the homogeneous Helmholtz equation:
\begin{equation}
\label{eq:helmholtz}
(\nabla^{2} + k^{2}) p(\bf{x}, \omega)=0,
\end{equation}
where $\omega$ is the wave frequency and $k=\frac{\omega}{c}$ is the wavenumber. 
When a specific boundary condition is given, its solution in 3D spherical coordinates can be decomposed as
\begin{equation}
\label{eq:solution}
p(r,\theta,\phi,\omega)=\sum_{l=0}^{\infty}h_{l}(kr) \sum_{m=-l}^{+l}c_{l}^{m}(\omega) Y_{l}^{m}(\theta, \phi),
\end{equation}
where $h_{l}(kr)$ is the Hankel function which encodes the radial pressure changes, $Y_{l}^{m}(\theta, \phi)$ is the spherical harmonics term which represents the angular part of the solution, and $c_{l}^{m}(\omega)$ is the spherical harmonics coefficients for frequency $\omega$. 
For a maximum order $L$, there are $(L+1)^2$ coefficients.  
Since the Hankel and the Spherical Harmonics functions are easy to evaluate, we only need to find the coefficients $c_{l}^{m}(\omega)$ to fully describe the acoustic scattering field. 

\subsection{Network Inference}
Our goal is to use an appropriate geometric representation for the underlying objects in the scene so that we can apply geometric deep learning methods to infer the acoustic scattering field from the object shape. 
It is also important that our approach should be able to handle highly dynamic scenes with general objects (i.e., no assumption on their convexity or genus or topology). 
Moreover, our approach should be able to generate accurate ASFs results for changing topologies. 
The input to our system  is the point cloud representation ($N\times 3$ matrix, $N$ being the number of 3D points) of an object. 
We represent each point and its local surface by a higher dimensional surface 
in the latent space formed by a surface encoder.  
 Based on Kolmogorov–Arnold representation theorem, we can represent the pressure field function as a function defined on a set of points ($N \times 3$), which is a multivariate continuous function composed of continuous functions of one variable encoded in a higher dimension. Moreover, borrowing the power of the \emph{De Finetti Theorem}, our pressure field function is further formulated as Eq.~\ref{th:definetti}. Please refer to \S~\ref{sec:learning} and \S~\ref{sec:newton} for more details of our design.
The desired output is the spherical harmonic coefficients vector up to order $L=3$ described in \S~\ref{sec:asf}. 
In practice, acoustic scattering corresponding to different frequencies can exhibit different distributions and the computational complexity increases as a cubic function of the frequency. 
Therefore, we train several networks for different frequency bands and in this paper we present the results for the frequency range $125Hz\sim 1000Hz$, which is sufficient for low-frequency sound propagation effects. In theory, our approach can extend to higher frequencies, though the cost of generating the training data would increase significantly.
We present a novel network architecture to compute these characteristics for arbitrary or deforming objects.

%% file: 4.tex
\section{Acoustic Scattering via Surface Encoding}
\label{sec:network_design}

In this section we introduce our shape Laplacian based point representation using surface encoders. For predicting ASF,  {the detailed geometric characteristics of object representation  play an essential role with respect to the simulation frequency.} This motivates our design in terms of using differential coordinates~\cite{sorkine2006differential}, which describe the discretization of the curvilinear integral around a given point (i.e., encodes the mean curvature). Moreover, we also provide the proof of its permutation invariant property via Newton's identities in the appendix,  {as that is important in the context of our sound propagation algorithm}. Finally, following De Finetti theorem~\cite{de_finetti}, we justify our design of the surface encoder and {show its benefits in terms of accurately computing the acoustic scattering functions for arbitrary objects}.

\subsection{{Local Surface Shape}}
\label{sec:local_shape}

Previous works on point cloud learning algorithms mostly focus on designing per-point operations~\cite{pointnet}, encoding per-point features to estimate continuous shape functions~\cite{xu2019disn}, or minimizing loss between a point normal vector and its connected vertices~\cite{liu2019learning}. We extend these ideas to compute the acoustic scattering functions.  

 {For each point in the input cloud and its neighborhood in the Euclidean space, we assume that it can form a piecewise smooth surface around the point.} 
Each point is encoded by the shared multi-layer perceptron (MLP)~\cite{mlp} and can be represented by a vector in the higher dimensional latent space $\mathcal{Z} \in \mathcal{R}^{128}$ (see Figure~\ref{fig:architecture}) 
 {Thus, a piecewise-linear approximation of the surface around a given point 
can be used to estimate the local surface shape, where the differential coordinate}~\cite{sorkine2006differential}
(i.e., $\delta-$ coordinates) of each vertex $v_i$ can be expressed as:
\begin{equation}
\label{eq:uniform_delta}
    \overrightarrow{\delta_i} = \frac{1}{d_i}\Sigma_{j \in \mathcal{N}(i)}{(\overrightarrow{v_i} - \overrightarrow{v_j})}.
\end{equation}
Here $\delta_i$ encapsulates the local surface shape, $\mathcal{N}(i)$ represents the $K$ nearest neighbors of vertex $v_i$ in the Euclidean space, and $d_i = |\mathcal{N}(i)|$ is the number of immediate neighbors of $v_i$.
To estimate the  mean curvature of the local surfaces formed by each point and its spatial neighbors, we use the  radial basis function (RBF) $\varphi(\cdot) = \exp^{-||\cdot||^2}$ to weigh each vector, rather than using the uniform weight shown in Equation~(\ref{eq:uniform_delta}). Since there are $N!$ permutations for a point cloud with $N$ vertices, every operation on point clouds should be permutation invariant (i.e., the permutation of input points should not change the output of our network).  {Therefore, we design an algorithm based on discrete laplacian and prove its  permutation invariance property in \S~\ref{sec:newton}. }

Our weight function is designed to be positive definite and symmetric for any choice of data coordinates.

\subsection{{Discrete Laplacian Based Surface Encoder}}
\label{sec:implicit_surface}
To encapsulate the local surface shape, each point $v_i$ is projected onto a higher dimensional space using MLPs, and the high-dimensional surface is defined on the latent space $z$ as shown on the top of Figure~\ref{fig:architecture}. 

Specifically, for one layer MLP, the representation is $z_i = relu(\overrightarrow{w} \cdot \overrightarrow{\delta_i} + b)$, where $\overrightarrow{w}$ and $b$ are learnable parameters in our network. To calculate the $\delta-$ coordinates of a point $v_i$ and the closed simple surface curve around it, its immediate neighbors in the Euclidean space (i.e., $\mathcal{N}(v_i)$) is used to evaluate Equation~(\ref{eq:delta_implicit}):

\begin{equation}
\label{eq:delta_implicit}
\begin{split}
    \overrightarrow{\delta^{*}_i}(v_i) =  \Sigma_{j \in \mathcal{N}_{(i)}}{\frac{\varphi(\overrightarrow{v_i} - \overrightarrow{v_j})(\overrightarrow{v_i} - \overrightarrow{v_j})}{\Sigma_{j \in \mathcal{N}_{(i)}}\varphi(\overrightarrow{v_i} - \overrightarrow{v_j})}}.
\end{split}
\end{equation}
In the latent space, the local surface shape is encoded as a high-dimensional surface where $\mathcal{N}(z_i)$ is evaluated. The direction of the differential coordinate vector, also defined in Equation~(\ref{eq:delta_implicit}), approximates the local normal direction. 


In order to compare the two $\delta-$ coordinate representations, we highlight the error in decibel (dB) between the pressure fields reconstructed from groundtruth spherical harmonics term and the predicted ones using different neural networks in Table~\ref{tab:delta_comparison}. We observe that $\delta^{*}-$ coordinates result in smaller errors. This indicates that our formulation provides a better approximation of ASFs.

\subsection{ {Neural Network Design}}
\label{sec:learning}
\begin{figure}[htbp]
  \centering
  \includegraphics[  trim = 0 150 0 40, clip,width = \linewidth]{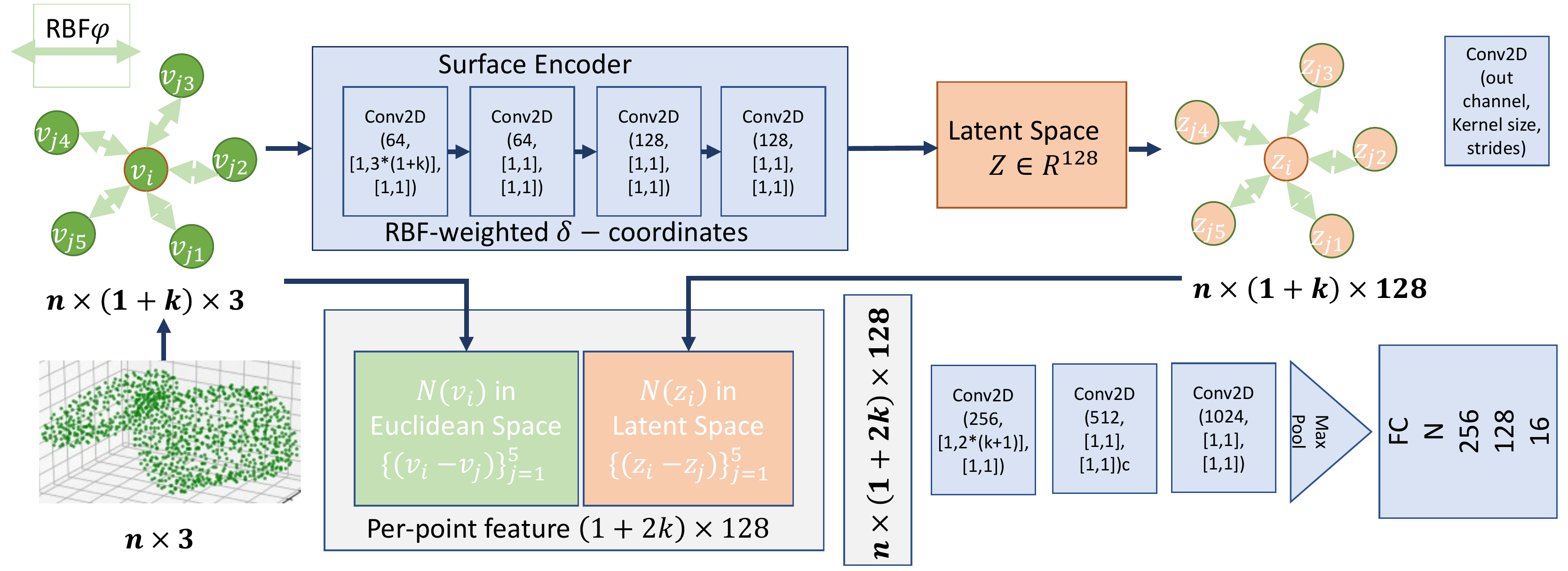}
  \caption{ 
  \textbf{Network architecture:} The input of our neural network is a $N \times 3$ point cloud ($N=1024$) and the output of our network is the spherical harmonic coefficients as a vector of length $16$ (up to $3^{rd}$ order spherical harmonics).
   {The piecewise smooth local shape around every point in the point cloud is encoded by virtue of the surface encoder, forming a high dimensional latent space in $\mathcal{R}^{128}$. The RBF-weighted $\delta-$ coordinates (Equation~(\ref{eq:delta_implicit})) where neighbors are found in the Euclidean space and the latent space are concatenated to represent the curvature feature, forming Equation~(\ref{eq:implicit_rep}). We leverage the geometric details by several Conv2D and MLP layers and output $16$ spherical harmonics coefficients $c_l^m$.} 
}
  \label{fig:architecture}
\end{figure}

Our neural network takes the point cloud as an $N \times 3$ input, where $N$ represents the number of points in the point cloud. The output is the spherical harmonic coefficients $c^m_l, -l \leq m \leq l, 0 \leq l \leq 3$, resulting in a vector of length $16$. The network is illustrated in Figure~\ref{fig:architecture}.

For each point $v_i$ in the $N \times 3$ point cloud, the discrete laplacian is evaluated on its immediate neighbors $\mathcal{N}(v_i)$ according to Equation~(\ref{eq:delta_implicit}).  
Then one convolutional layer and three MLP layers are used to encode the piecewise-smooth local shape around $v_i$ into a latent space $\mathcal{Z}\subset \mathcal{R}^{128}$. For each $z_i\in \mathcal{Z}$, the discrete laplacian Equation~(\ref{eq:delta_implicit}) is further evaluated where $\mathcal{N}(v_i)$ (i.e., neighbors of $v_i$ in the Euclidean space $\mathcal{R}^3$) and $\mathcal{N}(z_i)$ (i.e., neighbors of $z_i$ in the latent space $\mathcal{Z}^{128}$) scale the $\delta-$ coordinates $\mathcal{R}^{128}$, respectively. This results in a $2K \times 128$ matrix. The final representation of $v_i$, $feature(v_i)$ is: 
\begin{eqnarray}
    \bigg(  z_i^T,
    {\frac{\varphi(\overrightarrow{v_i} - \overrightarrow{v_1})(\overrightarrow{z_i} - \overrightarrow{z_1})}{\Sigma_{j \in \mathcal{N}(v_i)}\varphi(\overrightarrow{v_i} - \overrightarrow{v_j})}}^T 
    {\tiny, \dots 
    , }{\frac{\varphi(\overrightarrow{v_i} - \overrightarrow{v_K})(\overrightarrow{z_i} - \overrightarrow{z_K})}{\Sigma_{j \in \mathcal{N}(v_i)}\varphi(\overrightarrow{v_i} - \overrightarrow{v_j})}}^T, & & \nonumber \\
    {\frac{\varphi(\overrightarrow{z_i} - \overrightarrow{z_1})(\overrightarrow{z_i} - \overrightarrow{z_1})}{\Sigma_{j \in \mathcal{N}(z_i)}\varphi(\overrightarrow{z_i} - \overrightarrow{z_j})}}^T
    {\tiny, \dots 
    ,} {\frac{\varphi{(\overrightarrow{z_i} - \overrightarrow{z_K})}(\overrightarrow{z_i} - \overrightarrow{z_K})}{\Sigma_{j \in \mathcal{N}(z_i)}\varphi(\overrightarrow{z_i} - \overrightarrow{z_j})}}^T\bigg), & & \label{eq:implicit_rep}
\end{eqnarray}
where $K = 5$ is the number of the nearest neighbors and $128$ is the dimension of the latent space. Now we have the piecewise approximation of the local shape around $v_i$ as described in Equation~(\ref{eq:implicit_rep}), which is composed of the center point $z_i \in \mathcal{R}^{128}$ and the $2K \times 128$ weighted discrete laplacian matrix. The vectors are concatenated and convolved with kernels of size $[1,(1+2K)]$, 
followed by two MLP layers and four fully connected layers with \textit{tanh} activation functions since the value of $c^m_l$ ranges from $[-1, 1]$. The output of the network is the spherical harmonics coefficients $c^m_l$ and the loss function is the $\mathcal{L}2$ norm of the difference between the predicted $c^m_l$ and the ground truth calculated using \emph{FastBEM Acoustics solver}~\footnote{\url{https://www.fastbem.com/}}. 

\subsection{ {Discrete Laplacian and Permutation Invariance by De Finetti Theorem}}\label{sec:newton}
In this section we describe the permutation invariant property of the RBF-weighted discrete Laplacian Equation~(\ref{eq:implicit_rep}). 
We mainly discuss the function composed of vectors in the high dimensional latent space projected by the surface encoder shown on the top of Figure~\ref{fig:architecture}. 

Let $f_{i}(v_1, \dots, v_N) = \Sigma_{j\in N_(v_i)}\psi(v_i - v_j)$, where $\psi$ is a continuous function composed of MLP layers. 
by the nature of \emph{Kolmogorov–Arnold representation theorem} (or superposition theorem) and multi-layer neural networks~\cite{KURKOVA1992501}, every multivariate continuous function can be represented as a superposition of continuous functions of one variable. Thus, $f_{i}(v_1, \cdots, v_N)$ can be rewritten as: $\Phi\big( \Sigma_{j=1}^{N}\Psi(v_j)\big)$, where $\Psi : \mathcal{R}^3 \rightarrow \mathcal{R}^{2N+1}$ and $ \Phi : \mathcal{R}^{2N+1} \rightarrow \mathcal{R}^3$. In our work, we do not project $v$ onto $\mathcal{R}^{2N+1}$; instead, we encode $v$ as $z \in \mathcal{R}^{128}$.
First, we show that $f_i$ can be regarded as a  sum-of-power mapping defined by the coordinate functions. Second, we show that such mapping is \emph{injective} and has continuous inverse mapping. The domain of $f_i$ is a compact set, and thereby the image of $f_i$ is a compact set as well. Thus, $f_i$ is a homeomorphism between its domain and its image, and thereby is symmetric and homogeneous as per Newton's identities (details in Appendix~\ref{sec:app_newton}). Further, we demonstrate that $f_i$ is a \emph{permutation invariant continuous} function since it can be considered as compositions of sum-of-power mappings.

Consider $\Psi(v):=[1, v, v^2, \cdots, v^{M}], \Psi : \mathbb{R}^3 \rightarrow \mathbb{R}^{M + 1}$, where $M$ is the dimension of our surface encoder space. Let $E(v) = \Sigma_{j=1}^{M}\Psi(v_j)$, then
\begin{equation}
\begin{split}
    E(v) &= \big[M, \sum_{m=1}^{M} v_m, \sum_{m=1}^{M} v_m^2, \cdots, \sum_{m=1}^{M} v_m^M\big] \\
    &= \big[M, p_1(v_1, v_2, \cdots, v_M), p_2(v_1, v_2, \cdots, v_M), \\
    & \cdots, p_M(v_1, v_2, \cdots, v_M)\big],\\
    &=\big[ E_0(v_1, v_2, \cdots, v_M), E_1(v_1, v_2, \cdots, v_M)\\
    & \cdots,    E_M(v_1, v_2, \cdots, v_M) \big],\\
\end{split}
\end{equation}
where $p$ is the \textit{power-sums} (defined in Equation~(\ref{eq:power_sum}), Appendix~\ref{sec:proof}). We observe that $p$ is symmetric (i.e., does not change if we permute  $v_1, \cdots, v_M$) and homogeneous of degree $M$. $E_q$ is defined in Lemma~\ref{eq:injective}. 
According to Lemma~\ref{eq:injective}, each element in $E(v)$ is injective. According to Lemma~\ref{lemma:inverse_mapping} (Appendix~\ref{sec:proof}), $E(v)$ has inverse continuous mapping. Therefore, $E$ is a homeomorphism between $\mathcal{R}^3$ and $\mathcal{R}^{M+1}$. Since $\Phi$ is the composition of several continuous functions $\psi$, this implies the continuity of $\Phi$. 

\begin{lemma}
\label{eq:injective}
Let $\mathbb{X}$ = $\{(x_1, \cdots, x_M) \in [0,1]^M : x_1 \leq x_2 \leq \cdots \leq x_M \}$. 
The sum-of-power mapping $\mathbb{E} : \mathbb{X} \rightarrow \mathbb{R}^{M+1}$ defined by the coordinate functions :$    E_q(X) := \sum_{m=1}^{M}(x_m)^q, q = 0, \cdots, M $
is injective.
\end{lemma}
The proof of Lemma~\ref{eq:injective} is in Appendix $A$. We can conclude that there exists functions $\Phi$ and $\Psi$ such that $f_i$ is permutation invariant as long as $\Psi$ and $\Phi$ are continuous functions. 
Our idea is to approximate the \textit{scattering field} defined by a set of points (i.e., the closed simple surface curvature of a given point in the point cloud). Based on De Finetti theorem, we can derive the acoustic scattering function in Eq.~\ref{th:fscatter} (Appendix~\ref{sec:finetti}). We further highlight the benefit of our Discrete Laplacian based surface encoder by showing that it can be regarded as sum of power mappings. Thereby, Eq.~\ref{th:fscatter} holds as per its continuity.

%% file: 5.tex
\section{Evaluation and Application}

In this section, we describe how our network (shown in Fig. 2) is used to approximate the ASF of an object and use that ASF for interactive sound propagation.
\subsection{Data Generation and Training}
We need a sufficiently large dataset for training and evaluating our method. 
We sample 100,000 3D objects from the \emph{ABC Dataset}~\cite{Koch_2019_CVPR}. 
All mesh files are pre-processed to be randomly rotated, scaled between $1\sim 2m$ and centered.
The \emph{FastBEM Acoustics}
is used to generate accurate acoustic scattering fields. 
Specifically, we simulate a plane wave travelling to the $-x$ axis direction, and all objects are assumed to have a zero Neumann boundary condition to solve Equation~(\ref{eq:helmholtz}). 
While other boundary conditions are possible, we do not extensively study this variable in this work. 
After the acoustic scattering field has been solved at frequencies bands of $\{125,250,500,1000\}Hz$, we use the \emph{pyshtools}~\footnote{\url{https://shtools.oca.eu/shtools/public/index.html}} package to project the directional field to spherical harmonic coefficients $c_{l}^{m}$. 
By using a maximum order of $L=3$, we are able to maintain a relative projection error below $2\%$. 
The original object meshes are also converted to point clouds using furthest sampling for 1024 points. 
The data generation pipeline takes about 12 days using a desktop with 32 Intel(R) Xeon(R) Gold 5218 CPU cores. 

We split our dataset into training, validation and test set following a 8:1:1 ratio. Our neural network is trained on an NVIDIA GeForce RTX 2080 Ti GPU and takes less than 1ms per object for inference. 
The Adam optimizer is used and it decays exponentially with decay rate and decay step equal to $0.9$ and $10\times$ \#training examples, respectively.  

\subsection{ASF Computation: Results and Evaluation}

We evaluate our network performance on thousands of unseen objects from the \emph{ABC dataset}, perform ablation study on RBF-weighted $\delta-$ coordinates and implicit surface encoders, and show that our network is robust to Gaussian noise ($\sim N(0,0.05)$). 
We compare our results with PointNet~\cite{pointnet} and DGCNN~\cite{dgcnn} in Table~\mbox{\ref{tab:delta_comparison}} and Table~\ref{tab:robust}, respectively. We justify the design of our network with this ablation study, including the use of $\delta-$ coordinates, RBF-weighted function as well as the implicit surface encoder, as highlighted in  Figure~\mbox{\ref{fig:architecture}}. We use PointNet, where only per-point MLP layers are applied on each point in the point cloud, and DGCNN, which dynamically build graphs according to $K$ nearest neighbors, as the baseline. RBF-weighted $\delta-$ coordinates (as described in Equation~(\ref{eq:delta_implicit})), uniformly weighted $\delta-$ coordinates (described in Equation~(\ref{eq:uniform_delta})), and implicit surface encoder shown at the top of Figure~\mbox{\ref{fig:architecture}} are considered as subjects of ablation studies. We observe that our fine-grained geometric feature representation in Equation~(\ref{eq:implicit_rep}) results in larger reduction in dB error, especially at higher frequencies. This demonstrates improved accuracy of our approach over ~\cite{tang2020learning}. Moreover, our discrete Laplacian based representation is more robust to the Gaussian noise in the test set compared to DGCNN. We visualize some results from these networks in Figure~\ref{fig:ASF_result} and in Appendix. 
\begin{figure}[!htb]
    \centering
    \includegraphics[trim = 0 10 0 10, clip,width=0.9\linewidth]{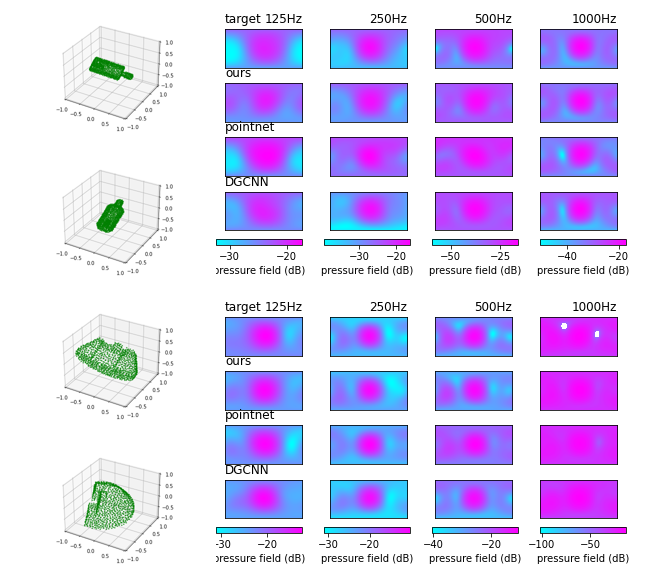}
    \caption{\textbf{ASF predictions and comparison:} We expand the predicted spherical harmonics coefficients onto a latitude-longitude (lat-long) map, representing the directional distribution of the scattering field. Here we show 2 sample point clouds from the unseen test set. Each point cloud is visualized from two viewing angles, followed by their groundtruth/target (Row 1), prediction using our method (Row 2), PointNet (Row 3), and DGCNN (Row 4). More results visualizations are in Appendix.}
    \label{fig:ASF_result}
\end{figure}

\begin{table}[htbp]
\begin{adjustbox}{max width = 1.1\linewidth}
    \centering
    \begin{tabular}{ccccccc}
    \toprule
        Methods &
        \multicolumn{4}{c}{Error in dB} &
        \multirow{2}{*}{\#Param}  &  \\
    \cline{2-5}
    \{Eq.~\ref{eq:delta_implicit}  | surface encoder\} & 125Hz & 250Hz & 500Hz & 1000Hz &  & \\
    
    \midrule
        {\{\xmark | \xmark \}} &
        3.49 & 3.56 & 3.71 & 4.23 & 30k & \\
         {\{\checkmark | \xmark \}} &
        3.38 & 3.41 & 3.57 & 4.47 &  30k & \\
        \{\xmark | \checkmark \} &
        3.28 & 3.38 & 3.52 & 3.85 & 30k & \\
    \{\checkmark | \checkmark \} (ours) &
     \textbf{2.98} & \textbf{3.06} & \textbf{3.22} & \textbf{3.76} & 34k & \\\hline
        {PointNet}~\cite{pointnet} &  3.66 & 3.44 & 3.43 & 4.16 & 80k & \\
    DGCNN~\cite{dgcnn} & 3.71 & 3.56 & 3.59 & 4.21& 103k & \\
    \bottomrule
    \end{tabular}
    \end{adjustbox}
    \caption{
    \textbf{Ablation study:}  
    In this evaluation, we compare the performance on uniform \mbox{$\delta-$} coordinate in Equation~(\mbox{\ref{eq:uniform_delta}}), weighted \mbox{$\delta-$} coordinates in Equation~(\mbox{\ref{eq:delta_implicit}}) and the surface encoder in Equation~(\mbox{\ref{eq:implicit_rep}}) on our test dataset (including \mbox{$10k$} objects) at frequency bands \{125Hz, 250Hz, 500Hz, 1000Hz\}. 
    The best result for each frequency is highlighted in {\bf bold} (lower error is better).  We alter between choosing Equation~(\mbox{\ref{eq:uniform_delta}}) and Equation~(\mbox{\ref{eq:delta_implicit}}) that results in four different combinations: Row 1 and 2, 3 and 4. 
    Next, we experiment the use of implicit surface encoder (Row 3 and 4). Our proposed network design (Row 4) shows superior performance in terms of ASF approximation for all four frequencies.}
    \label{tab:delta_comparison}
\end{table}

\begin{table}[htbp]
    \caption{\textbf{Robustness test}:  
    In this robustness test, we add i.i.d. noise $\sim N(0, 0.05)$ to every point in the point cloud in the test set. Note that DGCNN finds $K-$ neighbors at each iteration, similar to our approach. However, our algorithm is more robust to the noisy data input thanks to the design of the shape Laplacian based point representation. PointNet was used in~\protect\cite{tang2020learning} and our approach is more accurate at higher frequencies.
    }
    \label{tab:robust}
    \centering
    \begin{tabular}{ccccc}
    \toprule
    \multirow{2}{*}{Methods} &  \multicolumn{4}{c}{Error in dB}
    \\\cline{2-5}
    & 125Hz & 250Hz & 500Hz & 1000Hz\\
        \midrule
        PointNet & 5.88 & 5.50 & 6.02 & 8.37  \\
        DGCNN & 7.91 & 6.65 & 7.31 & 11.32 \\
        Ours & \textbf{5.20} & \textbf{5.09} & \textbf{4.70} & \textbf{5.51}  \\
    \bottomrule
    \end{tabular}
\end{table}

\subsection{Application to Interactive Sound  Propagation}

We use our ASF computation in a hybrid sound propagation system. Since our learning method can compute the ASF in less than $1$ms, our approach involves no precomputation and we are able to handle deforming objects or changing topologies. In particular, we combine the ASFs with state-of-the-art ray-tracing based acoustic simulator~\cite{schissler2017interactive}. 
In our hybrid system, the learned ASFs are used to simulate wave acoustics up to $1000Hz$ to compensate the low-frequency effects like sound diffraction and interference.
This choice of the simulation frequency of $1000Hz$ is similar to~\cite{raghuvanshi2018parametric} and~\cite{rungta2018diffraction}, as it can provide plausible sounds for games, virtual reality, and computer-aided designs. 

Conventional ray-tracing algorithm finds direct and indirect ray paths from a sound source to a receiver/listener. Next, a time delay and energy damping is calculated according to the path length and the order of reflections of each path. This information is used to compose an impulse response for real-time sound rendering~\cite{kuttruff2016room}. 



\begin{figure}[htbp]
  \centering
  \includegraphics[width=0.9\linewidth]{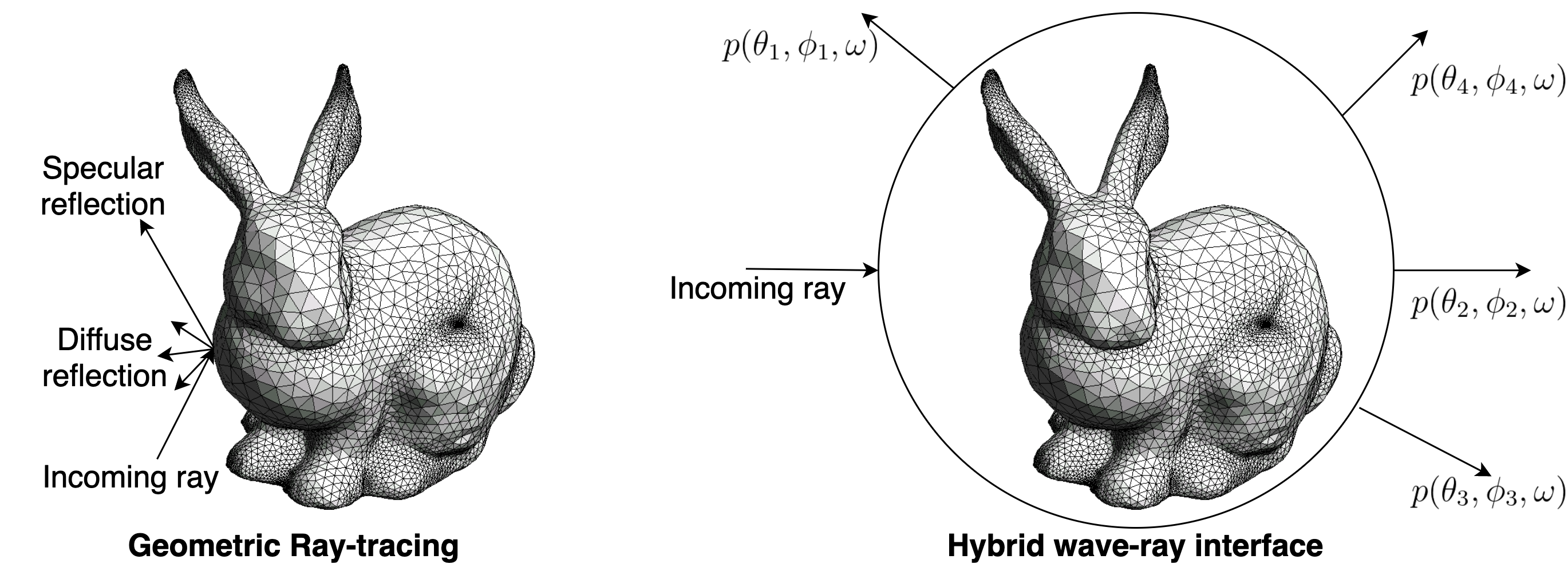}
  \caption{Pure geometric ray-tracing (left) vs our hybrid wave-ray coupling (right). Our learned ASF is used to approximate the scattering effects and compute multiple output directions.}
  \label{fig:raytracing}
\end{figure}

The core of the ray-tracing process is to sample a large number of reflection directions and their corresponding energy when a ray hits a surface. 
A pure geometric sound propagation system assumes specular and diffuse reflections based on high-frequency approximation (Figure~\ref{fig:raytracing} left). 
We use the ASF to approximate the diffraction and other wave-effects. 
Specifically, when an incoming ray hits a sound scatterer, we sample the outgoing directions among all directions and calculate the energy decay by evaluating the ASF at each direction $(\theta_i, \phi_i)$ for frequency $\omega$ (Figure~\ref{fig:raytracing} right). This hybrid wave-ray formulation is general enough to be used with most sound propagation engines.

\begin{figure}[!ht]
    \centering
    \subfloat[Floor: One static sound scatterer in open space. We compute the ASF and perform raytracing at 10.65ms/frame.]{\includegraphics[width=0.23\textwidth]{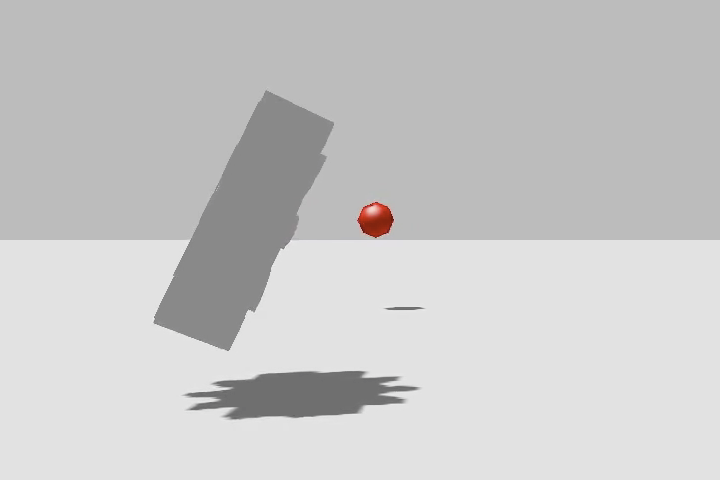}}\hfill
    \subfloat[Havana: Two moving walls in half-open space. We compute the ASF for each position of the wall for sound propagation at 6.78ms/frame.]{\includegraphics[width=0.23\textwidth]{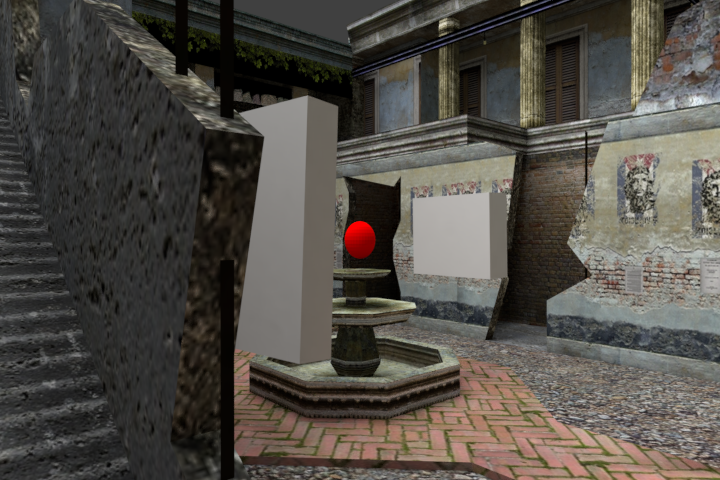}}\hfill
    \subfloat[Trinity: Six flying objects in a large indoor room. We compute the ASF for each object and use that for sound propagation at 12.95ms/frame.]{\includegraphics[width=0.23\textwidth]{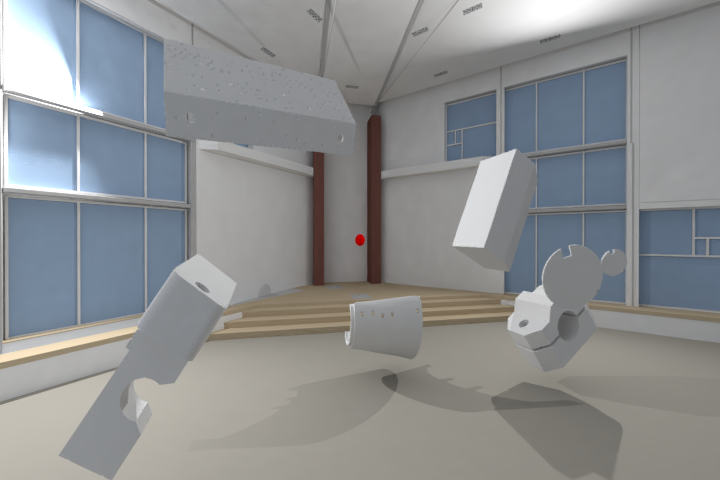}}\hfill
    \subfloat[Sibenik: Two disjoint revolving objects in a church. We compute their ASF and use them for sound propagation at 6.87ms/frame.]{\includegraphics[width=0.23\textwidth]{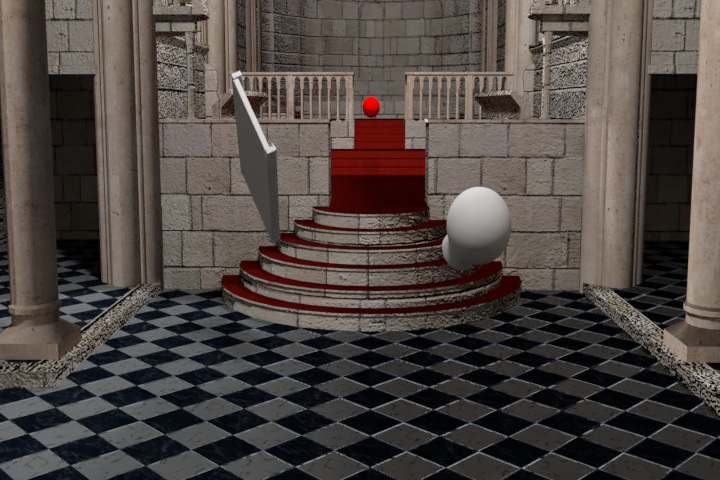}}
    \caption{Benchmark scenes used for audio-visual rendering in our supplemental video. These are dynamic scences where objects come in close proximity and change topologies. Our learning methods can compute accurate scattering fields and combine with ray tracing for interactive sound propagation.}
    \label{fig:benchmarks}
\end{figure}
We also evaluate the sound rendering in the four benchmark scenes with different complexity, as shown in Figure~\ref{fig:benchmarks}. We demonstrate that our ASFs can generate smooth and realistic sound renderings (in supplemental video).

%% file: 6.tex
\section{Conclusion and Limitations}
We present a novel geometric deep learning approach that infers the 3D acoustic scattering field induced by an object from its shape. 
Our algorithm uses discrete-laplacian based implicit function encoders to accurately capture the geometric properties of an object and achieves lower error compared with existing general point-based  learning methods. We use these scattered fields for interactive sound rendering and highlight their application in complex benchmarks.

Our current implementation is limited to sound-hard surfaces and a limited range of simulation frequencies, though these are not inherent limitations of our method. 
The simulation time needed to create a large training set can grow as a cubic function of the sound frequency. 
Incorporating more types of sound surfaces can further increase the training overhead of all geometric deep learning methods. 

%% file: appendix.tex
\section{Permutation Invariance}
\label{sec:proof}

In this section we provide the proof of permutation invariant property of the RBF-weighted discrete Laplacian Equation~(\ref{eq:implicit_rep}) thanks to Newton's identity. First of all, Sec.~\ref{sec:app_newton} serves as a recap of Newton's identity, which is applied in the proof of Lemma~\ref{eq:injective}. Lemma~\ref{lemma:inverse_mapping} shows that the \textit{sum-of-power mapping} defined in Lemma~\ref{eq:injective} has continuous inverse mapping. Both of the lemmas provide the background for the permutation invariance of our Discrete Laplacian in Equation~(\ref{eq:implicit_rep}). 
Furthermore, we demonstrate with the De Finetti Theorem that the acoustic scattering functions can be approximated using the underlying latent space. 
\subsection{Newton's Identities}
\label{sec:app_newton}
Newton's identities formulas, also known as \textbf{theory of equations}, build bridge between elementary symmetric polynomials and power sums. 
Consider a polynomial f of degree n with roots $x_1, \cdots, x_n$.
Assume that $s_0 = 1$, $f$ is monic. 

\begin{equation}
\label{eq:define}
\begin{split}
    f(x) &= s_0 x^n + \cdots + s_{n-1}x + s_n \\
        &= \prod_{i=1}^{n}(x - x_i) \\
    s_r &= (-1)^r \sum_{j_1 < \cdots < j_r} x_{j_1} \cdots x_{j_r} \\ 
    s'_r &= s_r \cdot (-1)^r \\
\end{split}
\end{equation}

Note that \textbf{$s_r$} and \textbf{$s'_r$} are \textbf{symmetric} (i.e. does not change if we permute $ x_1, \cdots , x_n $) and \textbf{homogeneous} of degree $r$.

These $s'_r$ polynomials, often called \textbf{elementary symmetric polynomials}, form a bases for all symmetric polynomials.

For example,
\begin{equation}
    \begin{split}
        s_1 &= (-1) * \big(x_1 + \cdots + x_n\big) \\
        s_2 &= x_1x_2 + x_1x_3 + x_1x_4 + \cdots \\
        s_3 &= (-1) * \big(x_1x_2x_3 + x_1x_2x_4 + x_1x_2x_5 + \cdots \big) \\
        & \vdots \\
        s_n &= (-1)^n * \big(x_1x_2x_3\cdots x_n\big) \\
    \end{split}
\end{equation}

Another family of symmetric polynomials are \textbf{power sums}, which form a basis for the space of all symmetric polynomials:
\begin{equation}
\label{eq:power_sum}
    p_r(x_1, \cdots, x_n) = x_1^r + \cdots + x_n^r
\end{equation}
The transition formulas between power sums and elementary symmetric polynomials are called \textbf{Newton's identities}.

\textbf{Lemma 4.1}
Let $\mathbb{X}$ = $\{(x_1, \cdots, x_M) \in [0,1]^M : x_1 \leq x_2 \leq \cdots \leq x_M \}$. 
The sum-of-power mapping $\mathbb{E} : \mathbb{X} \rightarrow \mathbb{R}^{M+1}$ defined by the coordinate functions :

\begin{equation}
    E_q(X) := \sum_{m=1}^{M}(x_m)^q, q = 0, \cdots, M
\end{equation}
is injective. 

\textbf{Proof: }{For some $u, v \in \mathbb{X}, E(u) = E(v)$, show $u = v$.}

Let $u,v$ be the roots of polynomial of degree $M$, we have:
\begin{equation}
    \begin{split}
        f_u(x) &= \prod_{m=1}^M(x - u_m) \\
        &= x^M + a_1x^{M-1} + \cdots + a_{M-1}x + a_M \\
        f_v(x) &= \prod_{m=1}^M(x - v_m) \\
        &= x^M + b_1x^{M-1} + \cdots + b_{M-1}x + b_M \\
    \end{split}
\end{equation}

$a, b$ are elementary symmetric polynomials as described in Equation~(\ref{eq:define}).
Elementary polynomials(i.e. $a,b$) can be expressed by power sums.
{\tiny
\begin{equation}
    \begin{split}
        a_M &= \frac{1}{M!} \begin{vmatrix}
        E_1(u) & 1 & 0 & 0 & \cdots & 0 \\
        E_2(u) & E_1(u) & 2 & 0 & \cdots & 0 \\
        E_{M-1}(u) & E_{M-2}(u) & E_{M-3}(u) & E_{M-4}(u)  & \cdots & M-1 \\
        E_{M}(u) & E_{M-1}(u) & E_{M-2}(u) & E_{M-3}(u) & \cdots & E_1(u) \\
        \end{vmatrix} \\
        b_M &= \frac{1}{M!} \begin{vmatrix}
        E_1(v) & 1 & 0 & 0 & \cdots & 0 \\
        E_2(v) & E_1(v) & 2 & 0 & \cdots & 0 \\
        E_{M-1}(v) & E_{M-2}(v) & E_{M-3}(v) & E_{M-4}(v)  & \cdots & M-1 \\
        E_{M}(v) & E_{M-1}(v) & E_{M-2}(v) & E_{M-3}(v) & \cdots & E_1(v) \\
        \end{vmatrix} \\
    \end{split}
\end{equation}
}
$E(u) = E(v)$ implies that $a = b$, and therefore $u=v$.

\begin{theorem}
\label{thm:homeomorphism}
The function $f : \mathbb{C}^M \rightarrow \mathbb{C}^M$, $f(c) = x^M + c_1x^{M-1} + \cdots + (-1)^{M-1}c_{M-1}x + (-1)^M c_M, c \in \mathbb{C}^M$, is homeomorphism~\cite{CURGUS200681}.
\end{theorem}

\begin{lemma}
\label{lemma:inverse_mapping}
The sum-of-power mapping defined in Lemma~\ref{eq:injective} has continuous inverse mapping.
\end{lemma}
The domain of $E$ is a compact set, and $E$ is a continuous function. Therefore, the image of $E$ is a compact set. 
From Lemma~\ref{eq:injective}, $a$ is a continuous function of the power sums $E$. 

\textbf{Goal:} Show continuity of inverse mapping of $E$. 

\textbf{Proof:} From Theorem~\ref{thm:homeomorphism}, the continuity of roots $u$ depends on the coefficients $a$. W.L.G, show continuity from $a$ to $u$. Due to the nature of homeomorphism, the mapping from $u$ to $a$ as well as its inverse mapping from $a$ to $u$ are both continuous.


\subsection{De Finetti Theorem and Implicit Surface Functions}\label{sec:finetti}
Our idea is to approximate the \textit{pressure field} defined by a set of points. 
And we borrow power from the following theorem~\cite{de_finetti}:
{\small
\begin{theorem}[De Finetti Theorem]
A sequence $(x_1, x_2, \dots, x_n)$ is \textit{infinitely exchangeable} iff
\begin{equation}
\label{th:definetti}
    p(x_1, \dots, x_n) = \int \prod_{i=1}^{n}p(x_i|\theta)P(d\theta),
\end{equation}
for some measure $P$ on $\theta$.
\end{theorem}

}
Thanks to the continuity nature of neural networks, we replace $P(d\theta)$ in Theorem~(\ref{th:definetti}) with a more general form $P(\theta)d\theta$ since we assume that the underlying latent layer is \textit{absolute continuous} according to Radon–Nikodym theorem (i.e., the distribution on $\theta$ has density.  {Note that some singular measures such as Dirac delta \textit{does not have density (Radon–Nikodym derivative)}. 
For example, each Radon measure (including Lebesgue measure , Haar measure, Dirac measure) can be decomposed into one absolutely continuous measure and one mutually singular measure.}). 
Then we derive that the acoustic scattering function of a given set of points $\mathcal{X}=\{x_1,x_2,...,x_n\}$ and the underlying latent space $\mathcal{Z}$ can be formulated in the form: 
\begin{equation}
\label{th:fscatter}
    f_{scatter}(\mathcal{X}) =\int dz \int \prod_{i=1}^{n}p(x_i|\theta)P(\theta | z )d\theta.
\end{equation}
 {Since our latent space vector $z$ can be regarded as compositions of sum-of-power mappings (Equation~(\ref{eq:power_sum}), Appendix~\ref{sec:proof}), the output pressure field of our network is permutation invariant.}

\begin{figure*}[b]
      \includegraphics[trim={0cm 0cm 0cm 0cm},clip,width=1.0\linewidth]{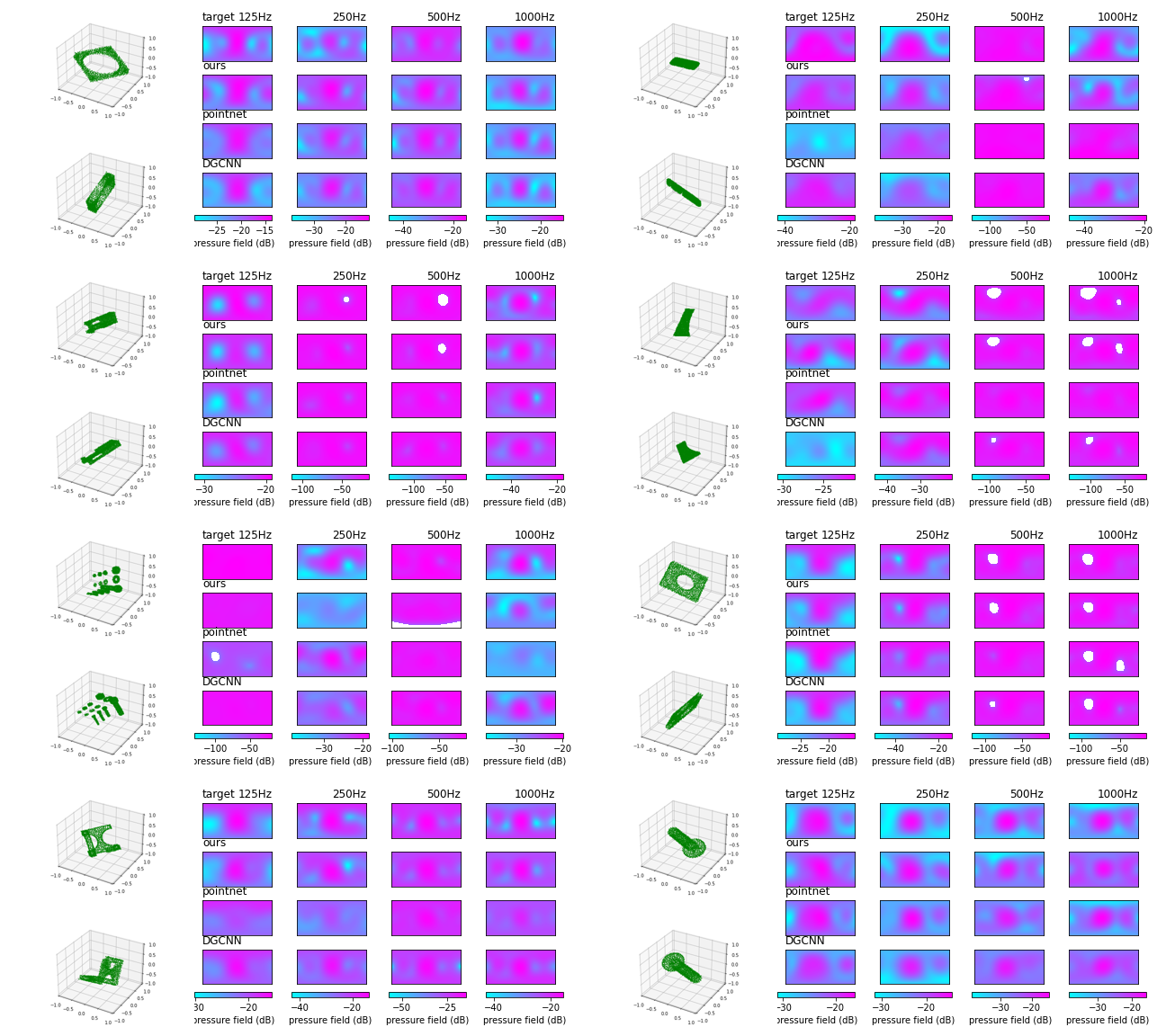}
      \caption{Additional acoustic scattering fields comparisons. Similar to Figure~\ref{fig:ASF_result}.}
\end{figure*}

\begin{figure*}[b]
  \includegraphics[trim={0cm 0cm 0cm 0cm},clip,width=1.0\linewidth]{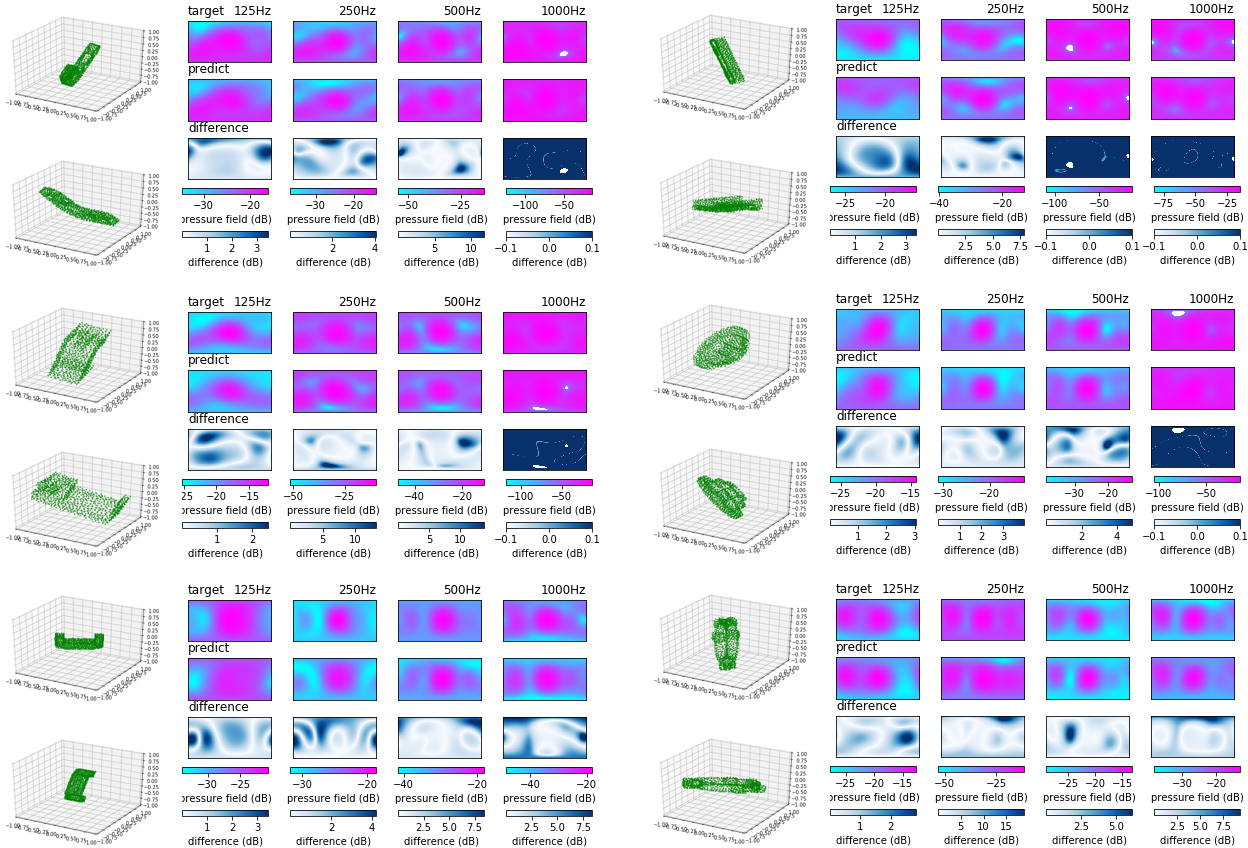}

  \caption{ASF prediction and groundtruth comparison: We expand the predicted spherical harmonics coefficients onto a latitude-longitude (lat-long) map, representing the directional distribution of the scattering field. We randomly choose 6 point clouds from the test dataset, which have not been seen during training. Each point cloud is visualized from two viewing angles, followed by their groundtruth/target, predicted, and error/difference lat-long maps in four frequency bands. We see a good match between our predicted fields and the groundtruth fields.
  }
  \label{fig:supp}
\end{figure*}

\begin{figure*}[b]
  \includegraphics[trim={0cm 0cm 0cm 0cm},clip,width=1.0\linewidth]{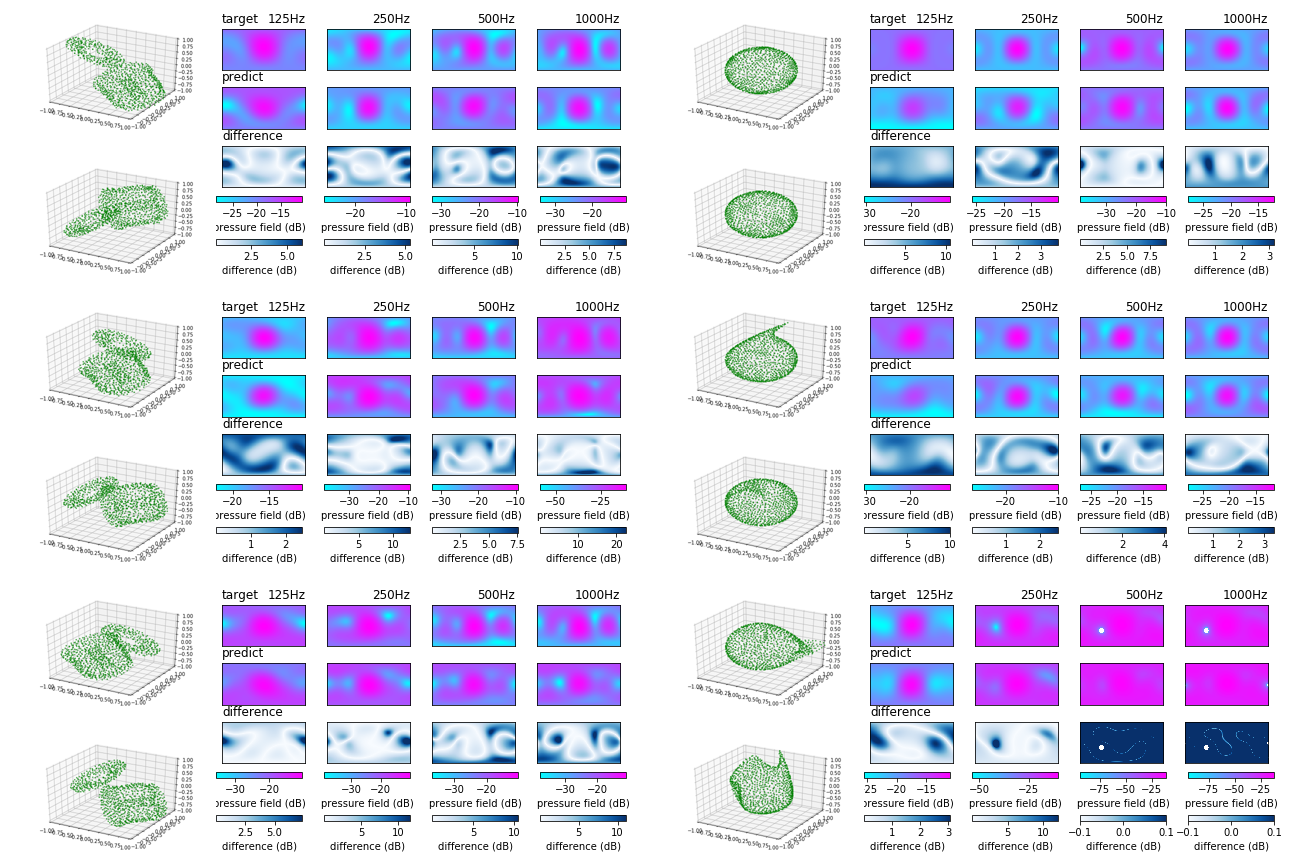}

  \caption{{\bf Moving and deforming objects:} The left column represents two objects moving close to each other. When the objects are very close, our approach treats them as one unified object and computes its point cloud representation. The right column shows different frames of a deforming sphere.
   The training dataset (from \emph{ABC Dataset}) does not contain deforming objects nor moving objects. Our algorithm generates  good approximations to ASFs for such dynamic objects, as we compare with the exact BEM solver.
  }
  \label{fig:supp_dynamic}
\end{figure*}